\documentclass[useAMS,usenatbib]{mn2e}

\usepackage{txfonts}
\usepackage[T1]{fontenc}
\usepackage[english]{babel}
\usepackage[latin1]{inputenc}
\usepackage{graphicx}

\newcommand{\eg}{e.g.,\,}
\newcommand{\ie}{i.e.,\,}

\newcommand{\be}{\begin{equation}}
\newcommand{\ee}{\end{equation}}
\newcommand{\bea}{\begin{eqnarray}}
\newcommand{\eea}{\end{eqnarray}}

\def\ifundefined#1{\expandafter\ifx\csname#1\endcsname\relax}
\def\la{\mathrel{\hbox{\rlap{\hbox{\lower4pt\hbox{$\sim$}}}\hbox{$<$}}}}
\def\ga{\mathrel{\hbox{\rlap{\hbox{\lower4pt\hbox{$\sim$}}}\hbox{$>$}}}}

\def\ang{\mbox{\AA}}

\title{Steps for Solving the Radiative Transfer Equation for Arbitrary Flows in Stationary
  Spacetimes}
\author[B.~Chen et al.]{B.~Chen$^1$\thanks{email:chen@nhn.ou.edu}, R.~Kantowski$^1$,
  E.~Baron$^{1,2}$, S.~Knop$^{1,3}$, P.~H.~Hauschildt$^3$\\
$^1$Homer L.~Dodge Dept.~of Physics and Astronomy, University of
Oklahoma, 440 West Brooks, Rm.~100, Norman, OK 73019, USA\\
$^2$Computational Research Division, Lawrence Berkeley
  National Laboratory, MS 50F-1650, 1 Cyclotron Rd, Berkeley, CA
  94720-8139 USA\\
$^3$Hamburger Sternwarte, Gojenbergsweg 112,
21029 Hamburg, Germany}

\begin{document}

\date{\today}

\pagerange{\pageref{firstpage}--\pageref{lastpage}} \pubyear{2002}

\maketitle

\label{firstpage}

\begin{abstract}
  We derive the radiative transfer equation for arbitrary stationary
  relativistic flows in stationary spacetimes, \ie for steady-state
  transfer problems.  We show how the standard
  characteristics method of solution developed by Mihalas and used
  throughout the radiative transfer community can be adapted to
  multi-dimensional applications with isotropic sources.
  Because the characteristics always coincide with geodesics
  and can always be specified by constants, direct integration of the
  characteristics derived from the transfer equation as commonly done
  in 1-D applications is not required. The characteristics are known
  for a specified metric from the geodesics. We give details in both
  flat and static spherically symmetric spacetimes. This work has
  direct application in 3-dimensional simulations of supernovae,
  gamma-ray bursts, and active galactic nuclei, as well as in modeling
  neutron star atmospheres.
\end{abstract}

\maketitle
\section{Introduction}\label{sec:intro}
The solution of the equation of radiative transfer in relativistic
flows is of considerable astrophysical interest. Steady-state and
dynamical solutions of the transfer equation are particularly
important for supernovae (SNe), gamma-ray bursts (GRB), and active
galactic nuclei (AGN). The general form of the general relativistic
transfer equation was derived by  \citet{lind66}, who also
derived the equation needed for neutrino transport in spherically
symmetric flows such as the core-collapse of massive stars. This work
was further extended in \citet{wilson71}, \citet{br85}, and
\citet{bmcv89}. In the 
stellar community the fully-special relativistic transfer equation was
derived and discussed by  \citet{mih80}.  General relativistic
versions of the transfer equation have been derived and discussed by
\citet{morita84,morita86}, \citet{schindblud89}, 
 \citet{zane96},   
 as well by 
\citet{Castor04}. 
\citet{mih80} proposed solving the transfer equation in the
``comoving'' frame using the method of characteristics. In 
this frame the momenta are measured by an
observer moving with the flow, and the spatial coordinates are those
of an inertial observer.  In the steady-state spherically symmetric
case the specific intensity $I_\nu$ is a function of only three
variables, the \emph{inertial frame} radius $r$, and two comoving
momentum coordinates, the energy $\varepsilon={\rm h}\nu$ and $\mu$
the cosine of the angle between the direction of the photon's momentum
and the radial direction.  Comoving frames are particularly useful
because they simplify the form of the the collision terms in the
transport equation. Mihalas made further use of the spherical symmetry
by treating the spatial and momentum angle variation of $I_\nu$
separately from the $\partial I_\nu/\partial \nu$ term. Somewhat
confusingly he plotted ``characteristic lines'', which have one
variable in real space and another in momentum space (see Fig. 1 of
\citet{mih80}).  These plots are curved lines and even though
practitioners know that they are working in a mixed frame, it is
common parlance to say ``the characteristics are curved''. 
Here we
re-emphasize that photons move along geodesics (straight lines in flat
spacetime) but demonstrate that for isotropic sources only one
momentum variable (the energy) must be
be comoving. We show that even in arbitrary 
3-dimensional flows one can choose parameters (coordinates) to label
geodesics which do not change along phase-space characteristics
(except for the affine parameter, or ``distance'' along the
characteristic itself). In addition we show that the change in
comoving wavelength along the characteristic can be handled by
standard finite difference techniques.  This procedure should simplify
the development of fully 3-D radiation transfer codes both in flat
space (applicable to variable stars, supernova and GRB spectra) and in
curved spacetime (applicable to neutron star atmospheres and AGN).

\citet{schindblud89} recognized that the momentum variables can be
chosen as constants and the transfer equation simplified in the
spherically symmetric case in the absence of a fluid flow. Although we
developed our formulation 
independently, our work is an extension of theirs to the
general 3-D case incorporating the effects of fluid flow. Their work
used the method of variable Eddington factors, whereas our method is a
characteristic based method and is specifically applicable to the case
of arbitrary fluid flow.

In Sections~\ref{sec:beq} and \ref{sec:rte} we introduce the Boltzmann
and radiative transfer equations and the relevant phase-space quantities. In
Sections \ref{sec:flat} and \ref{sec:sphere} we look at
characteristics in flat and spherically symmetric spacetimes
respectively.  In Section~\ref{sec:num} we discuss the logical steps
necessary to solve the steady state transfer equation for stationary
spacetimes and in Section~\ref{sec:disc} we give our concluding
remarks.
\section{The Boltzmann Equation for Photons}\label{sec:beq}
In this section we do not restrict spacetime, but we neglect
polarization effects. The Boltzmann Equation is an integro-differential
equation for the invariant photon distribution
function $F(x,p)$ on the photon's 7-dimensional phase-space $(x,p)$.
This can be thought of as the photon``on-shell" subspace of a full 8
dimensional  
particle phase-space.
The number of photons $\Delta N$ found by observer $u(x)$ in a small
6-element $\Delta V_x\Delta P$ of phase-space at $(x,p)$ is measured by 
the 6-form $\delta N$, \ie $\Delta N=\delta N(\Delta V_x,\Delta P)$
where
\be
\delta N\equiv F(x,p)\,\delta V_6,
\ee
and where  \citep[see][for details]{lind66}
\be
\delta\, V_6\equiv -(u(x)\cdot p)\,\delta\, V_x\,\delta P.
\ee
In the above, $u(x)$ is an arbitrary observer's unit 4-velocity at
spacetime point $x$, 
\be
-(u(x)\cdot p)=\mathrm{h}/\lambda
\label{lambda}
\ee
is the magnitude of the photon's 3-momentum as seen by observer
$u(x)$, $\delta\, V_x$ is the observer dependent 3-dimensional volume
element at $x$,  
and $\delta P$ is the
covariant volume element on the photons' 3-dimensional momentum space
at  $(x,p)$. Here, $\mathrm{h}$ is Planck's constant and $\lambda$ is the
wavelength measured by observer $u(x)$.

The collisionless Boltzmann equation
simply states that $F[x(\xi),p(\xi)]$ remains invariant (constant) 
along the Lagrangian flow of photons
in phase-space generated by their geodesic motion in spacetime.
Constancy of $F[x(\xi),p(\xi)]$ is a natural consequence of
Liouville's theorem, \ie $\delta V_6$
is invariant under this flow, and the constancy of $\Delta N$ due to
the absence of non-gravitational interactions of the
photons. Any lack of constancy of $\Delta N$ in a finite volume $\Delta V_6$ 
is accounted for by a collision term (\citet{oxenius} not withstanding).
To exhibit covariance, the Boltzmann equation with collisions, is often
written as a differential equation
\be
{dF\over d\xi}={dx^\alpha\over d\xi}{\partial F\over \partial x^\alpha}+
{dp^{\alpha}\over d\xi}{\partial F\over\partial p^{\alpha}}=
\left({dF\over d\xi}\right)_{coll},
\label{BE1}
\ee
with $F(x,p)$ explicitly given as a function of 8 variables (all 4
components of momentum are included but constrained by $p\cdot p=0$).
The collision term on the R.H.S. is a measure of the rate of change of
the number of photons $\Delta N$ in a $\Delta V_6$ transported along
the would-be paths of non-interacting photons in phase-space.

According to the geometrical optics approximation, photons  
travel on null spacetime
geodesics independently of their wavelengths. Affine parameters, $\xi$,
unique to each wavelength,
can be chosen
which generate the following orbits on
phase-space:
\bea {dx^\alpha\over d\xi}&=& p^\alpha,
\label{geodesic}\\
{dp^{\alpha}\over d\xi}&=&-\Gamma^{\alpha}_{\beta\gamma}p^\beta p^\gamma,
\label{geodesic-affine}
\eea
which reduces (\ref{BE1})  to
\be
p^\alpha {\partial F\over \partial x^\alpha}-
\Gamma^{\alpha}_{\beta\gamma}p^\beta p^\gamma{\partial F\over\partial p^{\alpha}}=
\left({dF\over d\xi}\right)_{coll}.
\label{BE2}
\ee
The R.H.S. is typically
separated into absorption and emission terms
\be \left({dF\over d\xi}\right)_{coll}=-fF+g,
\label{RHS}
\ee
where $f(x,p)$ and $g(x,p)$ are identified respectively with the
invariant absorptivity and emissivity \citep{morita86}. These
quantities implicitly depend on macroscopic properties of the
interacting medium such as temperature, pressure, and density.  The
geodesic equations (\ref{geodesic}) and (\ref{geodesic-affine}) are
equations for the characteristic curves of the integro-differential
PDE (\ref{BE2}).  These characteristic curves $(x(\xi),p(\xi))$ are
simply affinely parameterized spacetime geodesics, lifted to the
7-dimensional photon 
phase-space, which project back onto the null geodesics of
spacetime $x(\xi)$. 
It is important to understand that changing phase-space coordinates or
changing parameters for phase-space curves, 
doesn't alter these curves at all, only their description, \eg  
when mixed coordinates are used as in \citet{mih80} straight line
geodesics naturally appear curved. 

Logically, solving the Boltzmann equation is a two step process: first
solve the geodesic equations (\ref{geodesic}) and
(\ref{geodesic-affine}) for the required set of null geodesics and
second solve the Boltzmann equation (\ref{BE1}) with appropriate
boundary conditions. These two steps are combined in what is commonly
called the characteristics method where (\ref{BE2}) is solved by
changing the momentum variables to a comoving frame and the
characteristic parameter to a nonphysical distance. These phase-space
coordinate changes necessarily involve the fluid flow, and are unique
only in highly symmetric cases.  To tackle non-symmetric
flows/spacetimes, however, the two steps are best kept separate. First
solve the geodesic equations by using coordinates that are constant
along the geodesics (or by finding the geodesics in any coordinate
system and then transforming to constant coordinates) and second
proceed to solve equation (\ref{BE1}) in the form of the transfer
equation as given in the next section.

\section{The Radiation Transport Equation}\label{sec:rte}

The radiation transport equation is an integro-differential equation
for the specific intensity $I(x,p)$ which is equivalent to the
Boltzmann equation (\ref{BE1}) or (\ref{BE2}) for $F(x,p)$. Both are
functions on the photon's 7-dimensional phase-space $(x,p)$; however,
$I(x,p)$ depends on a choice of an observer at each point of spacetime
through \footnote{$I_\lambda(x,p)d\lambda dAd\Omega$ is the rate
  observer $u(x)$ detects energy crossing normal to his area $dA$ in
  the direction $p$, within his solid angle $d\Omega$, and in his
  waveband $d\lambda$.  Any locally-flat comoving reference frame at
  $x$ associated with the comoving observer $u(x)$ can be used to
  evaluate $dA$ and $d\Omega$.  These frames are arbitrary up to a
  rotation at each point $x$, however, actually defining one at every
  point $x$ isn't necessary.}  
\be I_\lambda(x,p)=-\frac{c^2}{\rm
  h}(u(x)\cdot p)^{5} F(x,p).
\label{I-F}
\ee
We have chosen to follow our group's convention and use $\lambda$ 
rather than $\nu$ as often appears in the literature; however, it is
straightforward  
to change between $I_\lambda$ and $I_\nu$ using  $\lambda
I_\lambda = \nu I_\nu$. 
Once  the observers are chosen, $I_\lambda(x,p)$ like $F(x,p)$, is a scalar.
Defining a set of observers is equivalent to giving a unit time-like
vector field on spacetime, $u(x)$, which appears in Eq.\,(\ref{I-F}).
Just as in equation (\ref{lambda}),  $-u(x)\cdot
p$ is equal to the
the photon's momentum as seen by observer $u(x)$.
If $u(x)$ describes the
material fluid with which the photons interact, $I_\lambda$ is called the
comoving specific intensity and $\lambda$
the comoving wavelength.

The transport equation for
$I_\lambda(x,p)$ is obtained from equations (\ref{BE1}) and
(\ref{RHS}) by substituting $I_\lambda(x,p)$ for $F(x,p)$ using Eq.\,(\ref{I-F})
\be {dI_\lambda \over d\xi}= -(\chi_\lambda{{\rm
    h}\over\lambda}+{5\over\lambda}{d\lambda\over
  d\xi})I_\lambda+\eta_\lambda{{\rm h}\over\lambda},
\label{I1}
\ee
where the observer dependent absorptivity $\chi_\lambda$ and emissivity
$\eta_\lambda$ are related to $f$ and $g$ by
\bea
\chi_\lambda &=&-\frac{f}{(u(x)\cdot p)}, \nonumber\\
\eta_\lambda &=&\left(\frac{{\rm c}^2}{\rm h}\right)\,(u(x)\cdot p)^4\,g.
\eea 
The other term ($\propto d\lambda/d\xi$) on the right in
Eq.\,(\ref{I1}) is present because the definition of $I_\lambda$,
Eq.\,(\ref{I-F}), explicitly depends on comoving $\lambda$.  If as is
customary we divide the extinction into two parts: ``true absorption''
$\kappa_\lambda$ and ``scattering'' $\sigma_\lambda$, then
$\chi_\lambda = \kappa_\lambda + \sigma_\lambda$.  
For a comoving
observer we will also assume the emissivity is given by thermal
emission (true absorption opacity $\kappa_\lambda$ times the Planck
function $B_\lambda$) and that scattering is elastic and isotropic.
This assumption is inherently required for our present formulation. It
is beyond the scope of this work to consider anisotropic or inelastic
scattering, but it is not entirely clear to us that the method cannot
be extended to the more general case.
For a comoving observer, $\chi_\lambda$ depends only on the magnitude
of the momentum and not its direction (given isotropic sources), and
consequently is a function 
of only $x$ and $u\cdot p$ in an arbitrary coordinate system. 

If the
energy density in the radiation field is written as $\epsilon_\lambda
= 4\pi J_\lambda/c$ (where $J_\lambda$ is the classic 0$^{\rm th}$
Eddington moment \citep{mihalas78sa}, and $\epsilon_\lambda$ is defined
in Eq.\,(\ref{energy})), the emissivity becomes 
\be 
\eta_\lambda =
\kappa_\lambda B_\lambda + \sigma_\lambda J_\lambda.
\label{eta-lambda}
\ee
When choosing coordinates $(x,p)$ on phase-space for the purpose
of evaluating $I_\lambda$ it is obvious that $\lambda$ itself should
be one of the choices because it simplifies the evaluation of
$\chi_\lambda$.  This is in fact the raison d'etre for using $\lambda$
as one of the momentum coordinates.  When attempting a numerical
solution, any dependence of $\chi_\lambda$ on direction requires the
use of a large number of angles in the kinematically favored forward
direction.  Since as one moves around the interaction region, the
forward direction changes, numerically resolving the variation of
$\chi_\lambda$ with direction can make the computational requirements
enormous.

 The reader should note that the right hand side of
Eq.~(\ref{I1}) differs slightly from 
the standard form of the non-relativistic static radiation
transfer equation because the affine parameter $\xi$ is not
a physical distance. As we discuss below it coincides
with a distance (up to a constant) in some spaces, \eg flat spacetime.

\section{Flat Space Simplifications}\label{sec:flat}
When solving Eq.\,(\ref{I1}) for the comoving intensity $I_\lambda(x,p)$ or
Eq.\,(\ref{BE1}) for the Boltzmann distribution function $F(x,p)$, the
dimension of phase-space can effectively be reduced if there are
common symmetries in spacetime, the interacting medium, and the
boundary conditions. This is because the transport equation has to be
solved on only one of each equivalent set of characteristics on the full
phase-space. For example if the spacetime, $u(x)$, and the boundary
conditions are stationary (\ie have a timelike symmetry), a time
dimension can be factored out and the required part of phase-space
reduces to 6 dimensions (3 space and 3 momentum). If the spacetime,
the flow, and the boundary conditions are stationary and spherically
symmetric, phase-space reduces to 3 dimensions (1 space and 2
momentum).
\subsection{Arbitrary Stationary Flows}\label{sec:arbitrary}
For flat space with a stationary flow and
static boundary conditions, 6 dimensions are required (3 space and
3 momentum). For space coordinates we choose the 3 Euclidean values
$\mathbf{r}$ of the flat spacetime inertial system in which the flow
is stationary
\be
u(x)=u(\mathbf{r})=\gamma(\mathbf{r})(1,\mbox{\boldmath$\beta$}(\mathbf{r}) ). 
\ee
For a boundary we choose a sphere of fixed radius $R$ surrounding the
origin.  We assume the emission and absorption coefficients vanish on
and beyond $r=R$.  On $r=R$ we assume
$I_\lambda(x,p)=0$ for photons with incoming directions, \ie with
$\mathbf{n}\cdot \mathbf{r}<0$ [see Eq.\,(\ref{geodesic-flat}) below], and  what we look for by
integrating the transfer equation is the outgoing intensity on $r=R$, \ie
we seek $I_\lambda(x,p)$ for
photons with $\mathbf{n}\cdot
\mathbf{r}>0$.  To make Eq.\,(\ref{I1}) as easy to
solve as possible, we choose two of the three momentum variables from
the direction cosines of the photon's direction in the inertial system
\be
\frac{dx}{d\xi}=\left(\frac{dt}{d\xi},\frac{d\mathbf{r}}{d\xi}\right)=
\frac{{\rm h}}{\lambda_\infty}\,(1,\mathbf{n})=\frac{{\rm h}}{\lambda_\infty}\,(1,n^x,n^y,n^z),
\label{geodesic-flat}
\ee
\eg we choose $(n^x,n^y)$. The subscript $``\infty"$ refers to wavelength as seen by inertial observers. 
If we were to use $\lambda_\infty$ as a 3$^{rd}$ coordinate,
all momentum coordinates would be constant, see
Eq.\,(\ref{geodesic-flat}), and the  
characteristic equations (\ref{geodesic}) and (\ref{geodesic-affine}) would be trivial. 
However, to accommodate the procedure used to solve Eq.\,(\ref{I1})
we must use the comoving wavelength $\lambda$ as the third momentum coordinate:
\be
\lambda=-\frac{\rm h}{(u(x)\cdot p)}=\frac{\lambda_\infty}
{\gamma(\mathbf{r})(1-\mathbf{n}\mathbf{\cdot}\mbox{\boldmath$\beta$}(\mathbf{r}))}.
\label{lambda-flat}
\ee

The comoving specific intensity
$I_{\lambda}(x,p) = I_\lambda(\mathbf{r},\mathbf{n})$
then depends on the spatial position $\mathbf{r}$, two direction
angles in $\mathbf{n}$, and the comoving wavelength $\lambda$;
however, the  
comoving absorption $\chi_{\lambda}$ and emission $\eta_\lambda$ coefficients
are independent of the direction angles.
The obvious
reason for choosing the first 5 of these 6 phase-space variables is
that their dependence on the affine parameter $\xi$ along any characteristic 
is easy to determine. The Euclidean position
$\mathbf{r}(\xi)$ is linear in $\xi$ and given by
\be
\mathbf{r}(\xi)=\mathbf{b}+s\ \mathbf{n},
\label{geodesic-flat2}
\ee
where
\be
s\equiv \left({{\rm h}\over\lambda_\infty} \xi\right).
\label{s}
\ee
The Euclidean distance $s$ is measured in the fixed inertial 
frame and $\mathbf{b}$ the impact vector
defined by $\mathbf{b}\cdot\mathbf{n}=0$ (see Fig. 1).
If $\lambda_\infty$ is written as a combination
of $\lambda$ and $\mathbf{r}$ using Eq.\,(\ref{lambda-flat}), 
the combination in parenthesis in Eq.\,(\ref{geodesic-flat2}) still
remains constant and the  
position coordinate part of a characteristic curve is still a straight line.
The photon's direction $\mathbf{n}$ remains constant and 
although $\lambda(\xi)$ is a complicated function of $\xi$, it is
given explicitly by 
substituting Eq.\,(\ref{geodesic-flat2}) into Eq.\,(\ref{lambda-flat}).

\begin{figure*}
\includegraphics{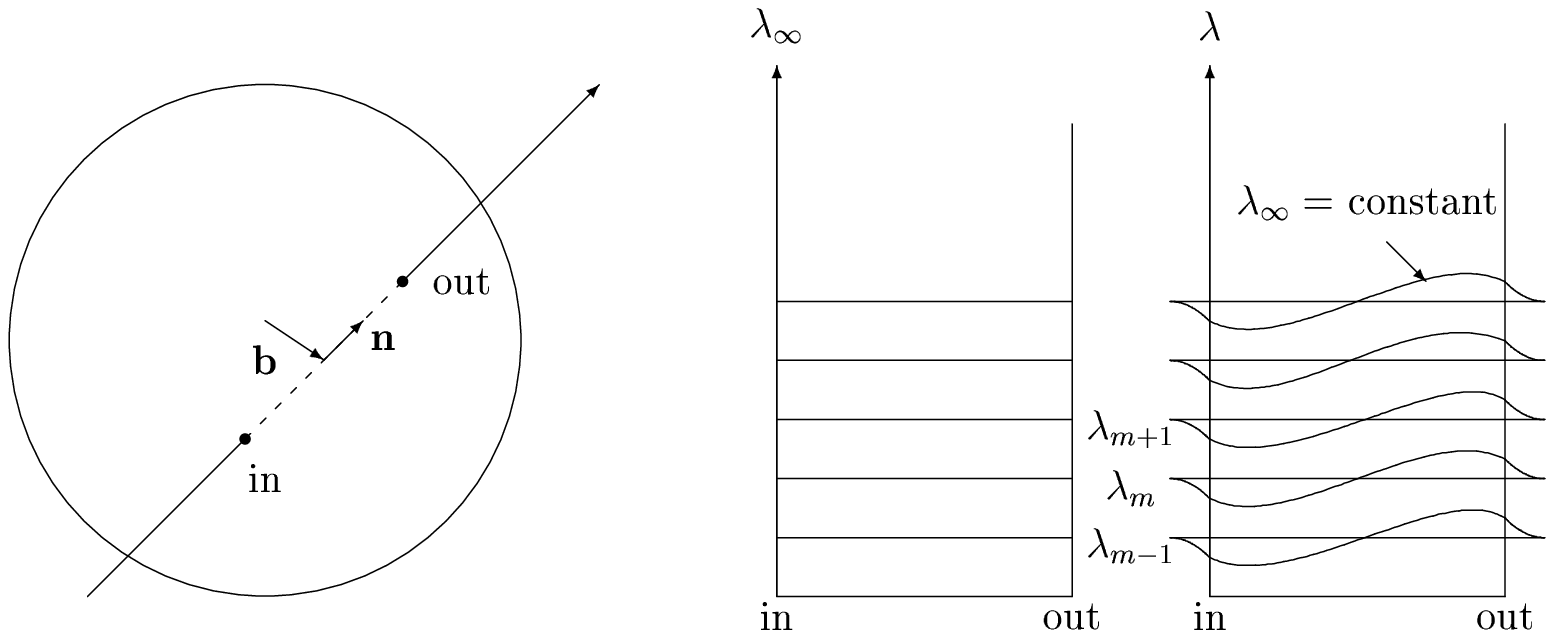}
\caption{On the left a single flat-space geodesic, Eq.\,(\ref{geodesic-flat2}), 
described by the impact vector $\mathbf{b}$ 
and tangent $\mathbf{n}$, enters and exits the boundary
of the interaction region $r=R$. On the right, the related characteristic
curves in phase-space corresponding to a discrete set of wavelengths
are also shown.
If $\lambda_\infty$ is used as the momentum coordinate
the characteristics of Eq.~(\ref{I2}) are all straight lines; however, if
the comoving $\lambda$ is used the characteristics deviate from being
straight but return to their $\lambda_\infty$ values beyond the boundary 
where the comoving fluid
coincides with the rest observers.  The characteristics of the differenced 
equation (\ref{I_m}) are defined only when $\lambda_m$ is constant.
 \label{fig:geos-flat}}
\end{figure*}
To logically connect $I_\lambda$ with distant observations we 
can (somewhat arbitrarily) smoothly distort the comoving fluid $u(x)$ to coincide with the
rest observers at some point beyond $r=R$ (see Fig. 1). Because $\chi_\lambda$ and 
$\eta_\lambda$ vanish in this domain, the fluid's effect on observations 
is sterile and $\lambda$ conveniently coincides with
$\lambda_\infty$. In most applications, the transfer from comoving to stationary observers
is done abruptly at the boundary and is implemented by a Lorentz 
boost. 
The reason for using the comoving wavelength as the 6$^{th}$
coordinate will become clearer when we alter the transport equation in
such a way as to have characteristic curves which keep $\lambda$ constant.
Rewriting Eq.\,(\ref{I1}) by explicitly separating out the $\lambda$
dependence we obtain
\be {\partial I_\lambda\over \partial \xi}\Biggr|_{\lambda}+
\left({d\lambda\over d\xi}\right) {\partial
  I_\lambda\over \partial\lambda}= -\left(\chi_\lambda{{\rm h}\over\lambda}+
\frac{5}{\lambda}{d\lambda\over d\xi}\right)I_\lambda+
\eta_\lambda{{\rm h}\over\lambda}.
\label{I2}
\ee
If Euclidean coordinates on flat space are used as in Eq.\,(\ref{geodesic-flat2}) 
the first term in Eq.\,(\ref{I2})  is related to the $\mathbf{r}$ dependence 
of $I_\lambda$ by
\be
{\partial I_\lambda \over \partial\xi}\Biggr|_{\lambda}=
{d\mathbf{r}\over d\xi}\cdot \nabla I_\lambda.
\label{characteristic-flat} 
\ee
The $d\lambda/d\xi$ part of the tangent to the characteristic is
computed from Eq.\,(\ref{lambda-flat}). If we would have chosen any 5
variables on phase-space, in addition to the comoving wavelength
$\lambda$, the transport equation would have been exactly of the form
Eq.\,(\ref{I2}).  As long as the new coordinates for phase-space are given as
functions of the original six, the characteristics in the new
coordinates are found by direct substitution.  The transfer equation
in the form of Eq.\,(\ref{I2}) says that $I_\lambda$ changes 
along a geodesic as usual because of
emission-absorption and its explicit dependence on $\lambda$ due to
its definition Eq.\,(\ref{I-F}); however, it also changes, when written as a function
of $\lambda$, because of its implicit
dependence on a comoving wavelength that changes ($\dot\lambda\ne 0$)
along the geodesic.  Another way to say the same thing is that when
geodesics on spacetime, are lifted to phase-space they are not
constrained to $\lambda=$ constant hypersurfaces (see Fig.\,1) and hence if
$\lambda$ is used as one of the coordinates, $I_\lambda$ changes
because $\lambda$ changes. 
 
\subsection{Radial Stationary Flows}\label{sec:radial}

In this section we try to clarify how our approach differs from the
classic approach of Mihalas. 
To make contact with what is commonly done
in the stationary spherically symmetric case where the flow is 
radial $\mbox{\boldmath$\beta$}=\beta(r)\hat{\mathbf{r}}$ and hence where phase-space
reduces to 2 independent dimensions beyond $\lambda$, we introduce 1
inertial coordinate $r$ 
and 1 comoving momentum coordinate $\mu$.
The wavelength and the radial coordinate can be evaluated from 
Eqs.\,(\ref{lambda-flat}), (\ref{geodesic-flat2}), and (\ref{s})
\bea
r(\xi)&=&\sqrt{b^2+s^2},\nonumber\\
\lambda(\xi)&=&\lambda_\infty \frac{\sqrt{1-\beta^2(r)}}{1-\beta(r)s/r}.
\label{r-lambda}
\eea
 The
coordinate $\mu$ is the cosine of the angle between the radial
direction and the direction of the photon in a frame instantaneously
moving with the radial flow at $\mathbf{r}$  
and is found by a radial Lorentz boost, 
\be
\mu(\xi)={\mathbf{n}\cdot\hat{\mathbf{r}}-\beta(r)\over
  1-\beta(r)\mathbf{n}\cdot\hat{\mathbf{r}} }=
\frac{ s/r-\beta(r) }{1-\beta(r) s/r }\ .
\label{mu}
\ee 
Equations (\ref{r-lambda}) and (\ref{mu}) are the integrated characteristic equation for 
the resulting transport equation for $I_{\lambda}(r,\mu)$,
\be
\left({dr\over d\xi}{\partial I_\lambda \over \partial r}+
{d\mu\over d\xi}{\partial I_\lambda\over \partial \mu}\right)+
{d\lambda\over d\xi}{\partial I_\lambda\over \partial\lambda}
=-\left(\chi_\lambda{{\rm h}\over\lambda}+
{5\over\lambda}{d\lambda\over d\xi}\right)I_\lambda+\eta_\lambda{{\rm h}\over\lambda},
\label{Mihalas3.1}
\ee
which is equivalent to
equation (3.1) of \citet{mih80} for $I_\nu$. Mihalas' parameter
$s_M$ is related to the standard affine parameter $\xi$ by
$ds_M=-(u\cdot p)d\xi=({\rm h}/\lambda) d\xi=(\lambda_\infty/\lambda) ds$, 
and equals the differential spatial distance
traveled by the photon as measured in the instantaneous local rest
frame of the comoving observer (and shouldn't be confused with
distance in any global frame). 
The ``comoving" distance parameter $s_M$
is the same for all photons with identical paths $\mathbf{r}(\xi)$ and
depends on the fluid velocity $u(x)$ they intercept, but not on their
individual wavelengths. When $s_M$ is used Eq. (\ref{Mihalas3.1}) becomes
\be
\left({dr\ \over ds_M}{\partial I_\lambda \over \partial r}+{d\mu\ \over ds_M}{\partial I_\lambda\over \partial \mu}\right)+
a(s_M,\lambda){\partial I_\lambda\over \partial\lambda}
=-\left(\chi_\lambda+{5\over{\rm h}}{d\lambda\over d\xi}\right)I_\lambda+\eta_\lambda,
\label{MihalasSN}
\ee
where $a(s_M,\lambda)$ is defined as
\be
a(s_M,\lambda)\equiv{\lambda\over {\rm h}}{d\lambda\over d\xi},
\ee 
and is related to Mihalas's $a_M(s_M,\nu)$  by
\be
a(s_M,\lambda)={\lambda^2\over {\rm h} }a_M(s_M,\nu)
\ee 
By using Eqs. (\ref{s}), (\ref{r-lambda}), and (\ref{mu}), $a(s_M,\lambda)$ is easily evaluated,
\begin{equation}
a(s_M,\lambda)=\gamma\left[(1-\mu^2){\beta\over r}+\gamma^2\mu(\mu+\beta){d\beta\over dr}\right]\lambda,
\label{a}
\end{equation}
as are the characteristic equations,
\begin{eqnarray}
{dr\ \over ds_M} &=& \gamma(\mu+\beta), \label{drds}\\ 
{d\mu\ \over ds_M}&=&\gamma(1-\mu^2)\left[{1+\mu\beta\over r}-
\gamma^2(\mu+\beta){d\beta\over dr}\right].
\label{dmuds}\end{eqnarray}
These are to be compared with Eqs. (3.4a), (3.4b) and (3.9) of \citet{mih80}.
We have arrived at the original transport equation obtained by \citet{mih80}, characteristics and all,
but do not propose solving Eq.~(\ref{MihalasSN})
However, if we were to now follow Mihalas, we would solve
the single equation, $ds_M=({\rm h}/\lambda) d\xi$, and obtain the integrated characteristics 
from (\ref{geodesic-flat2}) and (\ref{s}). By first finding the geodesics and secondly deciding on what 
variables to use on phase-space, we obtain the characteristics by substitution. 
For phase-space coordinates used in Eq. (\ref{MihalasSN}), the characteristic curves had 
three non-vanishing tangent vectors Eqs. (\ref{a}), (\ref{drds}), and (\ref{dmuds}).
The procedure we use is to choose as many coordinates on phase-space as possible that remain constant 
along geodesics (lifted to phase-space). For radial flows one constant coordinate is chosen from 
$\mathbf{b}$ and $\mathbf{n}$ \eg $b$, and the other two are $\xi$ itself and $\lambda$. 
The transport equation is of the form (\ref{I2})
with characteristics having only two non-vanishing tangents, 
\ie only two coordinates change along any characteristic, $\xi(\xi)=\xi$ and $\lambda(\xi)$ 
[see Eq. (\ref{r-lambda})]. Only one coordinate would change if we chose $\lambda_\infty$ rather than
$\lambda$. However, the differencing procedure described in \S 6 requires the comoving energy (\ie $\lambda$ for us)
to be used as one of the phase-space coordinates.   

 It is clear from the above discussion that if
an affine parameter such as $\xi$ is used, the characteristic curves are
completely determined and don't have to be constructed as integral curves
of their tangents. If a
non-affine parameter is used (\eg $s_M$) all that must be done is to
relate it to $\xi$ (\eg by inverting $s_M=-\int(u(x)\cdot p)d\xi$) along
the geodesic and substitute. For a numerical example of see \citet{bh04}.  
The affine parameter $\xi$ is changed to
optical depth $\tau_\lambda$ by a similar
substitution. A single reference wavelength such as 
$\lambda_{\mathrm{std}} = 5000$~\ang\
is usually chosen, making 
$d\tau_\mathrm{std} = \pm \chi_{\lambda_{\mathrm{std}}}ds_M$.\footnote{In
  classical plane-parallel and spherically symmetric radiative
  transfer $\tau$ is measured from the outside inward and the minus sign 
is appropriate.}
This choice greatly facilitates the generation of the spatial numerical grid.

\section{Static Spherically Symmetrical Spacetimes With Stationary
  Flows}\label{sec:sphere}
The relevant spacetime is
\be
ds^2=-e^{2\Phi(r)}c^2dt^2+e^{2\Lambda(r)}dr^2+r^2(d\theta^2+\sin^2\theta d\phi^2)\,,
\label{ss-metric}
\ee
which we assume is asymptotically flat \ie $\Lambda(\infty)=\Phi(\infty)=0$.
We assume boundary conditions similar to the flat case, \ie beyond $r=R$ the fluid 
becomes transparent ($\eta_\lambda=\chi_\lambda=0$) and 
that $I(x,p)$ vanishes for incoming photons.  
With a stationary, but otherwise arbitrary fluid flow, the effective dimension of phase-space 
reduces to 6 (3 space and 3 momentum). For this case we also distort $u(x)$ to coincide with static 
observers, 
\ie ($r,\theta,\phi$)=constants, at some finite $r$ value beyond $r=R$.

The metric (\ref{ss-metric}) has 4 Killing vectors, 1 time translation and 3 rotations, which 
aid in finding photon orbits. In Fig. 2 we show the 1-parameter family  of 
orbits confined to the $\theta=\pi/2$ plane 
and oriented symmetrically about $\phi=0$. They are labeled by the single impact parameter
$b$ and the wavelength $\lambda_\infty$ seen by a rest observer at $r=\infty$. 
The 3 non-vanishing momentum components are
\bea
cp^t={dct\over d\xi} &=&\frac{\rm h}{\lambda_\infty}\,e^{-2\Phi(r)},\\
p^\phi={d\phi_{\pi/2}\over d\xi}&=&\frac{\rm h}{\lambda_\infty}\,\frac{b\,e^{-\Phi(b)}}{r^2},\\
p^r&=&{dr\over d\xi}\\
&=&\pm\frac{\rm h}{\lambda_\infty}\frac{\,e^{-\Phi(b)-\Lambda(r)}}{r}
\,\sqrt{r^2e^{-2[\Phi(r)-\Phi(b)]}-b^2},\nonumber
\label{p^r}
\eea
from which the spacetime orbits are 
\bea
ct(r)&=& \pm\int^r_b
\frac{re^{\Lambda(r)+\Phi(b)-2\Phi(r)}\ dr\,}
{\sqrt{r^2e^{-2[\Phi(r)-\Phi(b)]}-b^2}},\\
\phi_{\pi/2}(r)&=&\pm\int^r_b\frac{\,b\,e^{\Lambda(r)}\ dr}{r\sqrt{r^2e^{-2[\Phi(r)-\Phi(b)]}-b^2}}.
\label{phi-r}
\eea
\begin{figure*}
\includegraphics{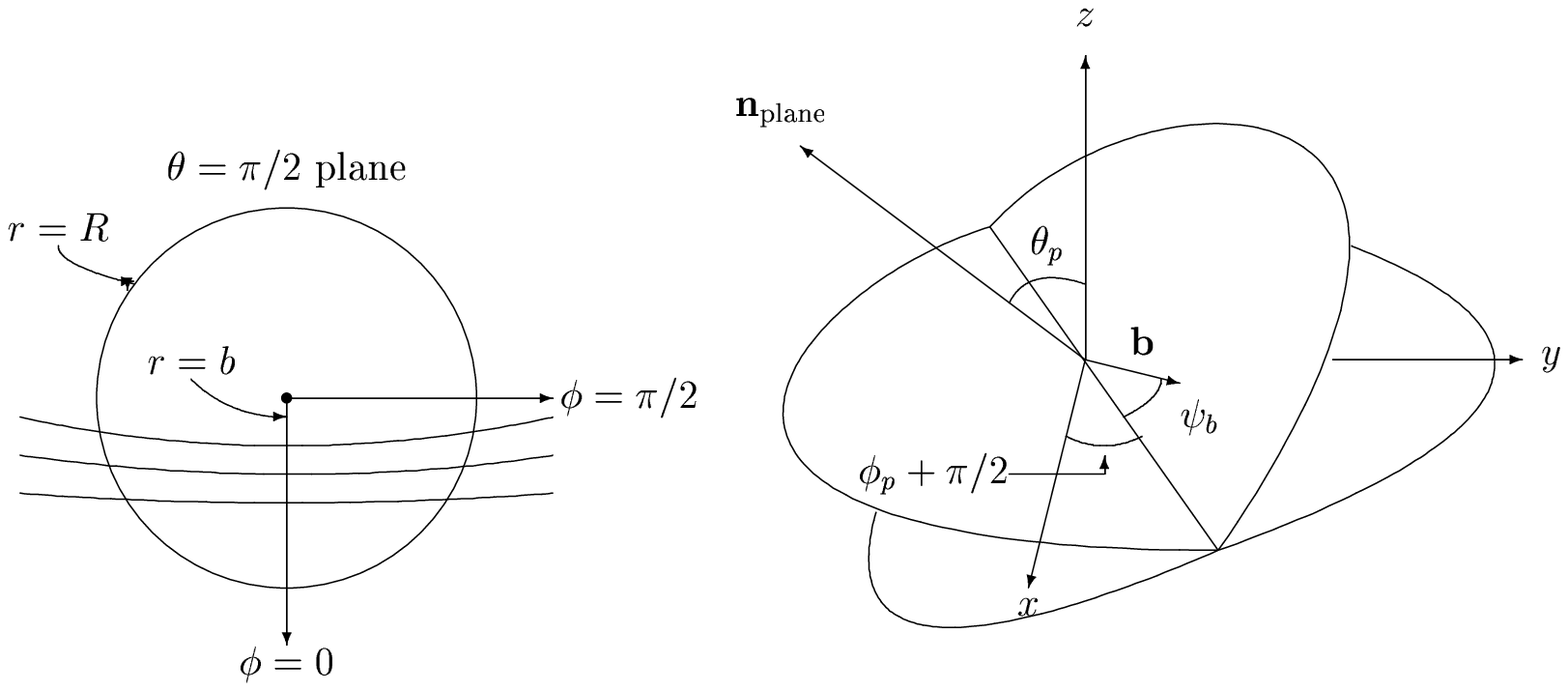}
\caption{On the left three null geodesics, which lie in the $\theta=\pi/2$ plane 
and are symmetric about $\phi=0$, 
are shown. On the right the active Euler rotations are shown that rotate the $\theta=\pi/2$ plane
and its geodesics into the plane whose normal is $\mathbf{n}_{\rm plane}$. The impact vector 
$b\,\hat{\mbox{\textbf{\i}}}$ is rotated into the 
$\mathbf{b}$ direction.
 \label{fig:geos-ss}}
\end{figure*}
All other photon orbits are obtained from these by rotations. 
According to Euler, any active rotation may be described 
using three rotations $R_z(\phi_p+\pi/2)R_x(\theta_p)R_z(\psi_b)$. 
For our purposes, 
$(\phi_p,\theta_p)$ specifies the direction of the normal to the photon's orbital plane 
$\mathbf{n}_{plane}$, and $\psi_b$ specifies the direction of the impact vector $\mathbf{b}$
within the orbit plane (see Fig. 2). These directions are space-like with no 
$t$ components and are constrained by $\mathbf{n}_{plane}\cdot\mathbf{b}=0$.
Because the rotated orbit is orthogonal to  $\mathbf{n}_{plane}$  
\be
\cos[\phi(r)-\phi_p]=-\cot\theta_p\,\cot\theta(r).
\label{orbit-plane}
\ee
The actual orbit in parametric form $(r,\theta(r),\phi(r))$ is given by 
\bea
\cos\theta(r)&=&\sin\theta_p\sin[\phi_{\pi/2}(r)+\psi_b],\label{rotated-orbit-a}\\
\tan\phi(r)&=&\frac{\cos\theta_p\tan[\phi_{\pi/2}(r)+\psi_b]-\cot\phi_p}
{1+\cos\theta_p\cot\phi_p\tan[\phi_{\pi/2}(r)+\psi_b]},
\label{rotated-orbit-b}
\eea
where $\phi_{\pi/2}(r)$ is given by Eq.\,(\ref{phi-r}).
Equation (\ref{orbit-plane}) can be used in place of Eq.\,(\ref{rotated-orbit-b}) if desired.
For an arbitrary stationary flow, the only net
symmetry is a time translation. 
The specific intensity depends on six variables
\eg $I_\lambda=I_\lambda(\mathbf{r},\mathbf{p})$. 
Instead of using the spherical polar angles and spherical polar momentum coordinates, 
we use the radius $r$, 
the comoving wavelength $\lambda$,
and four constants. It is convenient to choose the 4 constants from the set of 5 given by 
the impact vector $\mathbf{b}$ and the orbit plane normal $\mathbf{n}_{plane}$, \eg 
the impact parameter $b$ and the 3 angles $(\phi_p,\theta_p,\psi_b)$ described above. 
We now have $I_\lambda=I_\lambda(r,\mathbf{n}_{plane},\mathbf{b})$,
where the $\lambda$ dependence is implied.
We could have eliminated $\lambda$ in favor of the constant $\lambda_\infty$, however,
as with the flat-space case, resolution of the atomic lines and
accurately calculating the angular integral [see Eq.\,(\ref{energy})]
dictates that we use the 
comoving wavelength. 
The transfer equation is still equation (\ref{I2}),
but now
 \be
{\partial I_\lambda \over \partial
  \xi}\Biggr|_{\lambda}={dr\over d\xi}{\partial I_\lambda\over\partial r}\,,
\label{characteristic-ss} 
\ee 
and the single coordinate part of the characteristic to solve is Eq.\,(\ref{p^r}). 
The $\lambda(\xi)$ part is found by substituting into
\bea
{{\rm h}\over\lambda}&=&-u\cdot p={{\rm h}\over\lambda_\infty}\, \gamma e^{-\Phi(b)}\Biggl\{
e^{\Phi(b)-\Phi(r)}\mp \nonumber\\
&&\frac{\sqrt{r^2e^{-2[\Phi(r)-\Phi(b)]}-b^2}}{r}\beta^R\nonumber\\
&+&\frac{b}{r}\sin\theta_p\sin[\phi(r)-\phi_p]\,\beta^\Theta-
\frac{b}{r}\frac{\cos\theta_p}{\sin\theta(r)}\beta^\Phi
\Biggr\},
\label{lambda-curved}
\eea 
where the unit comoving 4-velocity $u(x)$ has been written in terms of
an orthonormal tetrad  
adapted to the static spherical polar coordinates of Eq.\,(\ref{ss-metric}),
\bea
u(x)&=&\gamma\left[\left(\frac{e^{-\Phi(r)}}{c}\frac{\partial\ \,}{\partial t}\right)\right. +\nonumber\\
&&\beta^R\left(e^{-\Lambda(r)}\frac{\partial\ \, }{\partial r}\right)+
\beta^\Theta\left(\frac{1}{r}\frac{\partial\ \, }{\partial
    \theta}\right)\nonumber\\
&+&\left.
\beta^\Phi\left(\frac{1}{r\sin\theta}\frac{\partial\ \, }{\partial \phi}\right)
\right]. 
\label{beta}
\eea
We assume that $u(x)$, and hence the $\beta^i(r,\theta,\phi)$ are given by 
prior hydrodynamical calculations.
Again as was illustrated Fig.~\ref{fig:geos-flat}, if $\lambda$ is
being used as a coordinate,  
it varies along a characteristic.

Because of the paucity of hydro calculations in GR one 
can use $\beta$'s that come from Newtonian gravity 
calculations  as a reasonable approximation as long as 
the gravity remains weak. The $\beta^i$ used in 
Eq.\,(\ref{beta})
should be Newtonian components relative to a static spherical
polar coordinate system. The GR metric associated with the Newtonian 
calculation is given by
\bea
e^{2\Lambda}&=&1/\left(1-\frac{2Gm(r)}{c^2r}-\frac{\Lambda_0}{3}r^2\right)\nonumber\\
\Phi(r)&=&\int^r_0e^{2\Lambda}\left(\frac{4\pi G P_r}{c^4}+
\frac{GM(r)}{c^2r^3}-\frac{\Lambda_0}{3}\right)dr,
\eea
here $\Lambda_0$ is the cosmological constant, $M(r)$ is approximated as
the Newtonian mass contained in radius $r$,  
and $P_r(r)$   is the
radial component of the pressure 
as seen by a static observer in the Newtonian hydro calculation.
For a numerical application of this section to the static Schwarzschild case see 
\citet{khbgr07}.
\section{Logical Steps for Numerical Integration}\label{sec:num}
The solution of the spherically symmetric transfer equation for radially
moving
flows has been discussed in detail by Hauschildt and Baron
\citep{phhs392,hbjcam99,hbmathgesel04,hb04}. In particular
\citet{hb04} showed how to stably difference the $\partial
I_\lambda/\partial 
\lambda$ term. This method will also work in the more general case discussed here. 
We briefly describe the approach in simple terms, for more
details see \citet{hb04}.
One first selects
a fixed set of comoving wavelengths, $\lambda_m$ at which to evaluate
the specific intensity $I_m$, and treats $\partial I_\lambda/\partial
\lambda$ as a difference, e.g., 
\be {\partial I_\lambda\over \partial
  \lambda}={I_m-I_{m-1}\over \lambda_m-\lambda_{m-1}}.  
\label{difference}
\ee 

This turns
the PDE, Eq.\,(\ref{I2}), into a discrete set of coupled ODEs with a single
differential variable $\xi$ for the set of $I_m$. The choice of the
set $\{\lambda_m\}$ is dictated by the variation of
$\chi_\lambda$ with $\lambda$. Rearranging, Eq.\,(\ref{I2}) 
becomes
\bea {dI_m\over d\xi}&+&\left[{\dot{\lambda}_m\over
    \lambda_m-\lambda_{m-1}}+{5\dot{\lambda}_m\over
    \lambda_m}+\chi_m{{\rm h}\over\lambda_m}\right]I_m\nonumber\\
&=&\eta_m{{\rm
    h}\over\lambda_m}+{\dot{\lambda}_m\over
  \lambda_m-\lambda_{m-1}}I_{m-1},
\label{I_m}
\eea
\citep[see][]{mih80} where $\dot\lambda_m$ is determined by
differentiating equations (\ref{lambda-flat}) or (\ref{lambda-curved})
with respect to the affine parameter $\xi$.  Even though phase-space
is still 6-dimensional we are now attempting to find $I_\lambda$ only
on a discrete set of lines corresponding to constant comoving
wavelengths $\lambda_m$.  For both flat and static spherically
symmetric spacetimes, the remaining continuous variables are 3
position and 2 momentum coordinates.  To simplify solving
Eq.\,(\ref{I_m}), coordinates should be adapted to the particular
spacetime symmetry and even then there is significant flexibility. For
flat space we chose $\mathbf{r}$ and 2 of the $\mathbf{n}$ components
making Eq.\,(\ref{geodesic-flat2}) the characteristics. In Fig. 1 a
spacetime geodesic of Eq.\,(\ref{geodesic-flat2}) is shown as it
enters and exits the boundary $r=R$. The related characteristic curves
in phase-space corresponding to a discrete set of wavelengths are also
shown. If $\lambda_\infty$ is used as the 3$^{\rm rd}$ momentum
coordinate, the characteristics of Eq.\,(\ref{I2}) are all straight
lines; however, if the comoving $\lambda$ is used the characteristics
deviate from being straight but return somewhere outside the
boundaries where the comoving fluid coincides with the rest observers.
The $\lambda$ part of the characteristics of the differenced transfer
equation (\ref{I_m}) has also changed, \ie comoving $\lambda_m$ now
remains constant.  The differencing term in Eq.\,(\ref{difference}) is
an approximation which attempts to account for the change from
$\lambda_\infty=$ constant curves to $\lambda=$ constant curves along
which $I_\lambda$ is propagated.  In practice the differencing
procedure used is more complicated \citep{hb04}.  When Eq.\,(\ref{I2})
is solved for $I_\lambda$ at an exiting wavelength, \eg $\lambda_m$,
its value depends on $I_\lambda$ values along its prior path for a
continuous spectrum of neighboring $\lambda$ values. However, when
Eq.\,(\ref{I_m}) is solved for $I_m$ at an exiting point, its value
depends on $I_{m'}$ values for only a discrete set of neighboring
wavelength $\lambda_{m'}$. The discrete coupling appears in the RHS of
Eq.\,(\ref{I_m}) in $\eta_m$ and explicitly as $I_{m-1}$.

The phase-space picture for the static spherically symmetrical spacetime 
is quite similar. We chose $r$ and 
4 constants from $\mathbf{n}_{plane}$ and $\mathbf{b}$ (see Fig. 2)
and the only non-constant component of the characteristic curves of Eq.\,(\ref{I_m}) 
is given by Eq.\,(\ref{p^r}). 

Often the source term is the more complicated part 
of the transfer equation. 
It contains various moments of the distribution function
depending on the assumed physics of the photon-matter interactions. 
A detailed discussion of the moment formalism for steady state
transfer can be found in
\citet{Thorne81}.
To balance the interaction between the radiation field and
the material, \ie to obtain energy-momentum conservation of the
radiation and matter, one must
adjust the material's parameters such as temperature. The opacity and
Planck function depend on the temperature of the material and thus
this temperature has to be consistent with the rate energy is being deposited by
the photon gas (which is not in thermal equilibrium with the matter).
One obtains this consistency through use of the energy-momentum tensor of
the photon gas. Per unit comoving 
wavelength it is
defined as the comoving solid angle
integral 
\bea T^{\alpha\beta}_\lambda(x)&\equiv&c\int \frac{dP}{d\lambda}p^\alpha p^\beta F(x,p)\nonumber\\ 
&=&\frac{1}{c}\,\int I_\lambda(x,p)(n^\alpha_u+u^\alpha)(n^\beta_u+u^\beta)d\Omega,
\label{stress1}
\eea 
where we have decomposed the photon 4-momentum into two parts, one
along the observer's 4-velocity $u$, and another perpendicular to it,
$u\cdot n_u=0$, by defining 
\be p=-(p\cdot u)(n_u+u).  
\label{n_u}\ee
Equation
(\ref{stress1}) can be decomposed into energy, momentum, and pressure
densities per unit wavelength as seen by $u(x)$, 
\bea
T^{\alpha\beta}_\lambda(x)&=&\epsilon_\lambda(x) u^\alpha(x) u^\beta(x)+\nonumber\\
&&{1\over c}\,{\it f}^\alpha_\lambda(x) u^\beta(x)+{1\over c}\,{\it f}^\beta_\lambda(x)
u^\alpha(x)+ {\it p}^{\alpha\beta}_\lambda(x), 
\eea
by defining 
\bea
\epsilon_\lambda(x) &\equiv& {1\over c}\int I_\lambda(x,p)d\Omega, 
\label{energy}\\
{\it f}^\alpha_\lambda(x) &\equiv&\int I_\lambda(x,p)n^\alpha_u d\Omega,
\label{momentum} \\
p^{\alpha\beta}_\lambda(x) &\equiv&{1\over c} \int
I_\lambda(x,p)n^\alpha_u n^\beta_u d\Omega, 
\label{pressure}
\eea 
where 
${\it f}^\alpha_\lambda u_\alpha=0$, $p^{\alpha\beta}_\lambda u_\beta=0$ and
$g_{\alpha\beta}p^{\alpha\beta}_\lambda=\epsilon_\lambda=4\pi J_\lambda/c$, 
see Eq.\,(\ref{eta-lambda}).
The rate per unit comoving wavelength per unit volume 
that 4-momentum (energy/c and momentum) 
is being 
transferred to the photons is 
\bea
T^{\alpha\beta}_{\lambda\ ;\>\beta}&=&\kappa_\lambda\left[-\epsilon_\lambda +
\frac{4\pi}{c} B_\lambda\right]u^\alpha-\frac{1}{c}\chi_\lambda
f^\alpha_\lambda\nonumber\\
&=& \frac{4\pi}{c} \kappa_\lambda\left[B_\lambda -
  J_\lambda\right]u^\alpha-\frac{1}{c}\chi_\lambda 
f^\alpha_\lambda.  
\label{eq:div_em_tensor} 
\eea
The
familiar statement of radiative equilibrium (total absorptions equals
total emissions in steady-state) for a comoving observer
is the vanishing of the integral of $u_\alpha T^{\alpha\beta}_{\lambda\ ;\>\beta}$
over all wavelengths.
Following  \citet{lind66} it is convenient to
define the particle (photon) flux 4-vector
\bea
N_{\lambda}^\alpha &=& \int  \frac{dP}{d\lambda}\, p^\alpha F(x,p),\nonumber\\
         &=& \frac{\lambda}{\rm c^2h} \int d\Omega\,
         I_\lambda(x,p) (n_u^\alpha + u^\alpha),\nonumber\\
         &=& \frac{\lambda}{\rm ch} \left[\epsilon_\lambda(x)
         u^\alpha(x) + {1\over c}\,{\it f}^\alpha_\lambda(x)\right],
\label{N}
\eea
in terms of which the rate per unit comoving wavelength that the photon number density
is changing due to
absorption and emission by sources can be
computed as
\be 
c\,N^\alpha_{\lambda\ ;\>\alpha}=
\frac{4\pi}{c} \kappa_\lambda \frac{\lambda}{\rm h}\left[ 
B_\lambda - J_\lambda\right].
\label{divN}
\ee
By inserting Eq.\,(\ref{N}) into the left hand side of Eq.\,(\ref{divN})
we obtain the identity
\be 
\left[ (\epsilon_\lambda(x)
         u^\alpha(x))_{;\>\alpha} + {1\over c}\,{\it
           f}^\alpha_\lambda(x)_{;\>\alpha}\right]
= \frac{4\pi}{c}\kappa_\lambda \left[ 
B_\lambda - J_\lambda \right].
\label{eq:div_flux}
\ee
Integrating Eq.\,(\ref{eq:div_flux}) over $\lambda$, together with 
radiative equilibrium from Eq.\,(\ref{eq:div_em_tensor}), leads to the
familiar result that the divergence of the flux is zero for a static fluid 
(\ie for a fluid where 
$u(x)\propto$ the timelike Killing field). 
Equations (\ref{eq:div_em_tensor}) and (\ref{eq:div_flux}) are
used to enforce the condition of radiative equilibrium  \citep[energy
conservation, for example see][]{hbjcam99}.
At depth, it is necessary
to use the vanishing of the LHS of Eq.\,(\ref{eq:div_flux}) 
when integrated over $\lambda$, rather than the RHS,
 because at high optical depth $J_\lambda$ is very close to
  $B_\lambda$ and both are very large. Numerically, it is difficult to
  obtain an accurate result from the subtraction of two large numbers.

The above expressions for the decomposition of $T^{\alpha\beta}_{\lambda}$
and $N_{\lambda}^\alpha$ are valid in any coordinate system and only
require knowledge of  
$u(x)$ in that coordinate system. Choosing a comoving frame for an
comoving  
observer in a non-symmetric spacetime essentially 
introduces an arbitrary rotation 
at every point in space. Fortunately it is not necessary to pick such a frame.
When evaluating the comoving solid angle integrals in
Equations (\ref{energy})--(\ref{pressure}) using stationary
coordinates one eliminates  
the comoving element $d\Omega$ in favor of the
stationary solid angle $d\Omega_0$ using 
\be
d\Omega=\left(\frac{u_0\cdot p}{u\cdot p}\right)^2d\Omega_0. 
\label{solid-angles}
\ee
Here $u_0$ is a unit vector pointing in the stationary frame's $t$ direction. 
For the flat space case 
$u_0\cdot p=-$h/$\lambda_\infty$, $u\cdot p$ is given by
Eq.\,(\ref{lambda-flat}), and  
\be
d\Omega={1-\beta^2\over(1-\hbox{\boldmath$\beta$}\cdot\mathbf{n})^2}d\Omega_0. 
\ee
For the spherically symmetric gravity field
\be
u_0\cdot p= -\frac{\rm h}{\lambda_\infty}e^{-\Phi(r)},
\ee
and $u\cdot p$ is given by Eq.\,(\ref{lambda-curved}).
To evaluate  $d\Omega$ we might have been tempted to introduce two 
comoving momentum variables  $\mu$, the $cosine$
of the comoving polar angle, and $\omega$, the comoving azimuthal
angle, as customary.
However, because of the arbitrariness of the flow 
evaluating these variables along a characteristic would have added two more 
complicated characteristics to solve and done little to aid integrating
Eq.\,(\ref{I_m}).

Solving the transfer equation for an arbitrary stationary spacetime is
similar to the above flat space and spherically symmetric examples.  A
stationary spacetime is one that has a timelike Killing vector.  If
the Killing vector is irrotational or equivalently hypersurface
orthogonal, the spacetime is called static as both examples
were. Stationary coordinates $(t,x^i)$ can always be adopted to the
Killing vector, \eg $K=\partial/\partial t$, which make the metric
components independent of $t$. If the space is static, all $dt\otimes
dx^i$ cross terms of the metric can additionally be made to vanish by
appropriately choosing a new $t$ coordinate, $t\rightarrow t+f(x^i)$.
When stationary coordinates are used, a fixed spatial boundary can be
drawn around the source, beyond which photons no longer interact with
the comoving fluid.  The boundary is 2-dimensional and can be somewhat
more complicated than the $r=R$ of the two examples. Incoming photons
can be labeled uniquely by 5 constants, 2 from the position they
strike the boundary, 2 from the impact orientation with which they
strike, and 1 from their wavelengths at $\infty$ (assuming the space
is asymptotically flat). The exact position of any photon within the
interaction region is then given by these 5 constants and the affine
parameter $\xi$ via the geodesic equations for the given geometry, and
the comoving wavelength is given via Eq.\,(\ref{lambda}). 
Finding coordinates which are constants is also the 
essence of Hamilton-Jacobi 
theory of classical particle mechanics. A family of 
canonical transformation is sought on phase-space which 
takes the particles in phase-space from their initial 
positions to their positions at time t.  The particles keep
their initial values and only time changes. 
 Solving 
Eq.\,(\ref{I_m}) for the general case then proceeds exactly as in the examples.  
In the two
examples we used the existing symmetry of the spacetime to choose
constants to label the photons rather than the boundary impact
coordinates and angles.  Impact constants could have been used; they
can be computed from the symmetry constants we actually used, \ie from
$\mathbf{b}$ and $\mathbf{n}$.  However, it is easier to stick with
symmetry constants when they exist.
\section{Discussion}\label{sec:disc}

Our formulation of the  radiative transfer equation in
terms of comoving wavelengths and stationary coordinates, 
and the recognition that the
momentum directions can be pre-chosen by constants is the fundamental
result of this paper. \citet{schindblud89} recognized this for the
case of purely static (no flow) transfer in spherical symmetry. Since
the directions of 
the geodesics may 
be chosen for example at the boundary, the solution of the full 3-D
radiative transfer problem in the presence of arbitrary hydrodynamic
flows is very similar to the purely static case (no flow) described
for example by 
\citet{hb06}. In that method space is divided up
into a 3-D rectangular grid and long characteristics are followed from the 
outer boundary through
the computational domain. The directions 
that the characteristics followed
were simply chosen by dividing up the angular space at the boundary into equal
parts. Since we have shown 
that the momentum directions may be chosen by constants in both the flat-space
and curved-space, the procedure of \citet{hb06} can be used
with only the modification that Eqs.\,(\ref{lambda-flat}) and
(\ref{solid-angles}) have to be evaluated once at each grid point.
Naturally, detailed numerical tests have to be performed to ensure
that the material properties such as density and composition are well
resolved, and that there are enough ``angles'' to resolve the
momentum-space variation of $I_\lambda$. This differs from more
classical methods in which the equations of the characteristics are
solved in phase-space, with all momentum coordinates co-moving.

If horizons are present in the spacetime, modification of the 
procedure outlined in this section may be necessary. 
However, for most astrophysical systems the emission of the
radiation and the radiative transfer occur in the accretion disk (or
in winds and jets) \ie outside of any spacetime horizon. 
The obvious case where one would like to calculate both photon and
neutrino transport in the presence of horizons, would be the formation
of a collapsar, see for example \citet{mw99}; however, this would
involve general relativistic MHD with radiative transfer, a feat that
is far beyond current computational capabilities.

Extending 1-D transfer calculations (\eg spherical symmetry with
radial flows) to 3-D applications with arbitrary flows is currently of
wide interest because of the desire to include the effects of rotation
and Rayleigh-Taylor instabilities into stellar wind models,
supernovae, and gamma-ray bursts; as well as the rapid improvement in
computing capabilities which makes the extension possible.  Some
confusion exists in which 1-D structures can be extended to 3-D and
which cannot. We have pointed out that the wavelength $\lambda$ (or
equivalently frequency) seen by a comoving observer $u(x)$ is one that
not only can be used, it is perhaps essential. Whereas a comoving
frame is useful in the 1-D case, it should not be used in 3-D. Not
that a comoving frame can't be defined for a comoving observer, it's
just that there are too many possibilities and no natural way to make
a unique choice between them. Consequently, as many as three
additional functions of position (\eg three angles at each point of
spacetime) will be unnecessarily introduced into the transfer equation by
introducing a comoving frame. 
The procedure we advocate for 
solving non-symmetric transfer problems is (1) start with the given
spacetime geodesics,  
(2) change to appropriate coordinates on phase-space, and (3) solve
the transfer  
using Eq.\,(34). 

We have concentrated on stationary flows in stationary spacetimes
(steady-state) applications, emphasizing the fact that, in spite of
how coordinates might be chosen, characteristics are really geodesics
extended to phase-space as described by \citet{lind66} and
\citet{ehlers71a}. We
have also shown that by choosing an appropriate set of parameters
(coordinates) to label the geodesics, characteristics can essentially
become straight lines.  When one of the coordinates is chosen as
comoving wavelength $\lambda$, only that particular coordinate changes
non-linearly. This slight complication is rewarded by a resulting
simplification in the absorption and emission terms on the
``collision" side of the transfer equation. Additionally the use of
$\lambda$ allows one to convert the transfer equation from an
integro-PDE to a system of integro-ODE's by a differencing procedure
which restricts comoving wavelengths to a discrete set
$\{\lambda_m\}$. These wavelengths remain constant along the
characteristics of the now differenced transfer equation and greatly
simplifies constructing the formal solution using long or short 
characteristic methods.

Other authors have recognized the current need for adapting 1-D
methods of solution to 3-D problems in radiative transport.  
\citet{CLM05} derive the relativistic equation of transfer in
flat spacetime, with similar goals to this paper; however, their
approach is somewhat geared to the discrete ordinates matrix
exponential method of numerical solution, or to generalized variable
Eddington factor (moment) methods.

\citet{broderick05} also recognized the advantage of using
constant momentum variables, which he attempts to define through the
use of Fermi-Walker transported tetrads. Even though his use of these
tetrads isn't clear, parallel transport itself results in constant
momentum components. If the spacetime is curved, parallel transporting
along every geodesic results at an infinite number of tetrads at every
point.  And, if scattering exists, the relation between these tetrads
must be ascertained before the emissivity integral [see
Eqs.\,(\ref{eta-lambda}) and (\ref{energy})] can be evaluated.

In conclusion, we have presented a workable outline of solving 
the radiative transport equation  for many 3-D steady-state problems.

\section*{Acknowledgments}
This work was supported in part  by NASA grants
NAG5-3505 and NAG5-12127, and NSF grants AST-0204771 and AST-0307323, 
PHH was 
supported in part by the P\^ole Scientifique de Mod\'elisation
Num\'erique at ENS-Lyon. 


\bsp

\label{lastpage}

\end{document}